\documentclass[a4paper,11pt]{article}
\pdfoutput=1 

\usepackage{jcappub} 

\usepackage[T1]{fontenc} 

\usepackage[english]{babel}

\usepackage[]{hyperref}
\usepackage[nottoc]{tocbibind}
\usepackage{amsmath}
\usepackage{amssymb}
\usepackage{textcomp}
\usepackage{array,multirow}
\usepackage{slashed}
\usepackage{tabularx}
\usepackage{graphicx}
\usepackage{subcaption}

\def\bea{\begin{eqnarray}}
\def\eea{\end{eqnarray}}
 \def\be{\begin{equation}}
\def\ee{\end{equation}}
\newcommand{\bSigma}{\overline{\Sigma}}
\newcommand{\bsigma}{\overline{\sigma}}

\title{\boldmath Primordial black holes from D-parity breaking in SO(10) grand unified theory}

\author[a,1]{Sasmita Mishra,\note{Corresponding author.}}

\affiliation[a]{National Institute of Technology Rourkela, Sundargarh, Odisha, 
India, $769 008$}
\author[b]{Urjit A. Yajnik}

\affiliation[b]{Indian Institute of Technology Bombay, Powai, Mumbai, India, 
$400 076$}

\emailAdd{mishras@nitrkl.ac.in}
\emailAdd{yajnik@iitb.ac.in}

\abstract{The growing evidence of gravitational waves from binary black hole mergers has renewed the interest in study of primordial black holes (PBH). Here we study a mechanism for the formation of PBH from collapse of 
pseudo-topological domain walls which form out of equilibrium during inflation and then collapse post inflation. We apply the study to domain wall formation due to $D$-parity embedded in a supersymmetric grand unified theory (GUT) based on $SO(10)$ and compare the abundance of resulting PBH with the existing constraints. 
Thus the macroscopic relics can then be used to constrain or rule out a GUT,  or demand a refinement of the theory of PBH formation in this class of GUTs.}

\begin{document}
\maketitle
\flushbottom
\section{Introduction}
\label{sec:intro}

The LIGO and Virgo collaborations have confirmed the direct detection of gravitational waves 
\cite{LIGOScientific:2016aoc, LIGOScientific:2016sjg} and binary black hole 
mergers. The detection has also confirmed the existence of a population of stellar-mass black 
holes with masses $\sim (10 - 100)M_{\odot}$. In the catalogs the Advanced LIGO and 
Advanced Virgo gravitational wave detectors 
\cite{LIGOScientific:2018mvr,LIGOScientific:2020ibl} along with KAGRA detector  
\cite{LIGOScientific:2021djp} have reported  candidates
consistent with the coalescence of binary black holes.
This has revived the interest in the origin of massive black holes \cite{Volonteri:2021sfo}, in particular
in the study of primordial black hole (PBH) \cite{Caldwell:2022qsj} \cite{Barack:2018yly} \cite{Carr:2018rid} 
\cite{Sathyaprakash:2009xs} while the possibility of any of them being a member of the observed collapsing  
black hole binaries  is an open question \cite{Caldwell:2022qsj}. 
The PBHs are interesting as they could in principle be small enough for Hawking 
radiation to be important \cite{Hawking:1974rv}. The black holes radiate 
thermally and evaporate with a time scale $\tau = 10^{64}(M/M_{\odot})^3$yr.
So the PBHs with mass $<  10^{15}$g would have completely evaporated by the 
present epoch. Whereas the PBHs with mass $\sim 10^{15}$ g would be evaporating 
today contributing to the cosmological $\gamma-$ ray background, Galactic 
$\gamma-$ ray background, $\gamma-$ ray bursts, radio bursts or some of them 
might have clustered within Galactic halo. 
The PBHs with mass $> 10^{15}$ g would still survive today to be detectable by 
their gravitational effects. Such PBHs could also act as non-baryonic dark 
matter candidates owing to their origin in radiation dominated era. They could 
also act as seeds
of supermassive black holes in the centres of galaxies. They may
play an important role in the generation of cosmological structures.
The PBH formation scenarios rely on the collapse of inhomogeneities\cite{Carr:1975qj} 
which has been more recently treated in \cite{Ananda:2006af}, 
while PBH formation from the collapse of  topological defects has also been proposed (see for instance \cite{Jenkins:2020ctp}\cite{Gelmini:2022nim}). 
In this paper we study the formation of PBHs from the collapse of topological pseudo-defect domain walls that form 
in the course of  spontaneous gauge symmetry breaking phase transition.

Domain walls are topological defects that appear in theories with broken
discrete symmetry. Early studies of $SO(10)$ grand unified theory (GUT) in the 
context of cosmology \cite{Kibble:1982dd} identified the possibility of unstable 
domain wall formation in certain realisations of the model.
This situation generalises to non-equilibrium conditions in the early Universe when
some scalar fields develop transitory condensates, but falling into one of several
degenerate minima of the thermal effective potential. The vacua are related by some of the broken generators
of the group, however the barrier separating them is finite and there need be no topological obstruction
to transition from one vacuum to the other. Yet, the dynamical effects combined with causal horizons 
in the early Universe can give rise to domain walls. Such defects were studied and dubbed 
Topological Pseudo-Defects in \cite{Garg:2018trf}. As discussed in \cite{Kibble:1982dd}, the 
collapse of such pseudo-defects can result in PBH, and this would be an interesting signature of the GUT visible in the 
present Universe. For the case of SUSY $SO(10)$ pseudo-defects, there is indeed a danger
that some of them persist for energetic though not topological reasons. These can 
in principle degrade by the processes noted by Preskill and Vilenkin \cite{Preskill:1992ck}.
But long lived defects at the GUT scale can interfere with inflation and ruin the
observed isotropy of Cosmic Microwave Background (CMB) by the Planck satellite collaboration \cite{Planck:2018vyg}. 
Thus a mechanism involving additional ultraviolet  operators for their destabilisation was proposed in \cite{Banerjee:2018hlp}.
We explore this setup here further for its potential to originate the PBH.

We consider the non-equilibrium scenario along the lines considered in \cite{Rubin:2000dq,Rubin:2001yw,Khlopov:2004sc}. 
Accordingly, Bunch-Davis fluctuations cause a scalar
field to evolve, alternating between two or more of its degenerate minima, giving rise 
to pseudo-defect domain walls (DW). The pseudo-defects at this stage carry relatively
small localised energy density due to the local minima being shallow.
Over time the relevant wavelength  of the scalar field, and the sizes of the domains grow large 
enough to exit the de Sitter horizon. The pseudo-defect structure remains imprinted 
on this fluctuation. Later, after reheating, the  wavelengths
re-enter the horizon and the DW re-enter, with large surface tension due to the remaining 
medium having cooled off substantially. They then start to shrink because of the surface tension.
If the collapse continues till the entire  energy of such closed wall gets concentrated 
within its gravitational  radius then the PBH can form.
No conflict arises with standard cosmology, as the pseudo-defects  
either become PBH,  being topologically unstable, or degrade by processes mentioned previously. 

The interesting possibility  is that the same  condensate which gives rise to pseudo-defects 
also acts as the inflaton.  For Minimal Supersymmetric $SO(10)$ GUT (MSGUT) a scenario with an embedded inflaton 
has been  considered in a variety of contexts \cite{Jeannerot:1995yn,Kyae:2005vg,Fukuyama:2008dv,
Aulakh:2012st,Garg:2015mra,Leontaris:2016jty,Ellis:2016ipm}. 
However, for the considerations in this paper we will leave this 
possibility open and consider the occurrence of the pseudo-defects in conjunction 
with a generic inflation, even if the inflaton can be identified with scalar or gauge 
fields \cite{Maleknejad:2011jw} within the same GUT. Further, several of the above
mentioned scenarios rely on gauge singlet superfields. Thus we leave embedding
of the inflaton within the GUT to later, and focus on the wall collapse scenario of PBH formation.

The plan of the paper is as follows. In section \ref{sec:png} we discuss a generic model of formation of PBHs under non-equilibrium conditions.  
In section \ref{sec:so10} and its subsections we discuss  generic DW formation due to the structure of $SO(10)$ 
followed specifically by the role of $D$-parity in MSGUT $SO(10)$ model and apply the  non-equilibrium formalism 
to  obtain our main results. Section \ref{sec:concl} contains the summary and conclusions. In appendix \ref{sec:appA} we summarise the
observational constraints on pre-galactic black hole abundances against which we have compared our results.

 \section{The Non-equilibrium initial conditions}
 \label{sec:png}
In this section we discuss a mechanism for appearance of DWs in a 
theory of a complex scalar field and subsequent PBH formation in an inflationary
universe as considered in 
\cite{Rubin:2000dq,Rubin:2001yw,Khlopov:2004sc}. Although first discussed in the context of
axions, the whole scenario is fortuitously applicable to $SO(10)$ due to the 
discrete parity flipped local minima available in the latter.
Consider the pseudo-Nambu-Goldstone bosons (PNG) arising from symmetry breaking
of a complex scalar field. Its potential is chosen to be 
Eq.(\ref{eq:png-pot})   
  \begin{equation}
   V(\varphi) = \lambda (|\varphi|^2 -f^2/2)^2.
   \label{eq:png-pot}
  \end{equation}
signalling the breakdown of its $U(1)$ symmetry at a scale $f$.
The minima of the potential in this case are degenerate, with possible vacuum expectation values 
$\varphi = \frac{f}{\sqrt{2}}{\rm exp}(i \phi/f)$. Further, it is assumed that the residual axionic degree of freedom 
$\phi$ develops a periodic potential, as for example due to instanton effects,
\begin{equation}
\delta V = \Lambda^4 (1-\cos\theta).
\label{eq:deltaV}
\end{equation}
where $\theta = \phi/f$.
In this dynamics of two successive second order phase transitions it is assumed that
$f >> \Lambda$ and also,
$\Lambda \ll H$, where $H$ is the Hubble parameter during inflation. 

At the first stage the spontaneous breaking $U(1)$ symmetry fills the
Universe with global $U(1)$ strings. In the second stage of symmetry breaking 
which  happens at the post-inflation, the phase $\theta$ of the complex scalar field acquires minima
at the points $\theta_{\rm min} = 0, \pm 2\pi, \pm 4\pi..$ and the field acquires mass
$m_0 = 2f/\Lambda^2$ but also the extended configurations develop kinks as argued next.
The field  $\theta$ obeys the  standard equation of motion,
\begin{equation}
 \ddot{\theta}-\nabla^2\theta +3 H\dot{\theta} +\frac{d \delta V}{d \theta} =0.
 \label{eq:eom}
\end{equation}
This equation permits two interesting regimes. At the scale of the field theory, i.e., length scales  $\ll H^{-1}$, 
we can ignore the $H\dot{\theta}$ term and we get 
 kink-like or DW solutions where the field interpolates between two adjacent vacua.
The  domain wall solution perpendicular to x-axis 
  between two vacua $\phi_1 =0$ and $\phi_2 = 2 \pi f$
  is given by,
  \begin{equation}
 \phi_{\rm wall} = 4~ f ~{\rm arctan} \left({\rm exp}\left(\frac{x-x_0}{d}\right)\right)
 \end{equation}
where $x_0$ is the position of the centre of the wall and  the width of the wall is set by $d\approx f/2\Lambda^2$. 
The surface energy density of the wall is then taken to be $\sigma \simeq 8 f \Lambda^2$.
At early stages while $H$ dominates we can ignore
the axion potential of \eqref{eq:deltaV}. During this time however, the fluctuations of
$\theta$ in regions separated by causal horizon can differ randomly to the extent of 
\begin{equation}
  \delta \theta = \frac{H}{2\,\pi\, f}.
  \label{eq:deltatheta}
 \end{equation}
This gives rise to two crucial steps on which the present scenario relies :
\begin{itemize}
\item
Such fluctuations have wavelengths ranging in size from the particle horizon during 
inflation $H^{-1}$ to the inflation horizon $e^{(N_{max}-N)} H^{-1}$, where $N$ is the number of 
e-folds remaining before the end of inflation. 
The fluctuations continue to evolve during inflation, with the phase $\theta$ making 
quantum jumps of $\delta \theta$ during every e-fold 
and in every space volume with characteristic size of the order $H^{-1}$. 
When  the wavelength of the fluctuation becomes larger than the $H^{-1}$ the average 
amplitude of this fluctuation freezes out, with the differences \eqref{eq:deltatheta} 
imprinted on them. 
\item
Subsequent to inflation and reheating, the Universe begins to cool and the local
minima of the potential \eqref{eq:deltaV} begin to have significance. The fluctuations
subside and the large scales begin to re-enter the Friedmann-Lem{\^a}itres-Robertson-Walker (FLRW) horizon. The Hubble scale
now drops rapidly as $1/2t$, and $\delta V$ signals the formation of kink-like or 
DW solutions where the field interpolates between two adjacent vacua.
As mentioned earlier, the surface energy density of such walls is $\sigma \simeq 8 f 
\Lambda^2$. If the energy stored $E$ in the closed DWs 
fall within its gravitational radius $r_g = 2 E/M^2_{\rm Pl}$, they could collapse 
to the black holes due to their surface tension. 
 \end{itemize}
 
We consider time evolution as a count down in terms of $N$ the number of e-folds \textit{remaining} before
the end of inflation, beginning with value $N_{\rm max}$. For the generic initial value of the classical 
field $\theta$ whose initial value is $\theta_{N_{\rm max}}$, one may assume $\theta_{N_{\rm in}} <\pi$. 
The presence of the de Sitter-like 
phase  gives rise to  characteristic  quantum fluctuations. 
The initial domain containing phase $\theta_{N_{\rm max}}$ increases its volume 
in $e^3$ times after one e-fold and contains $e^3$ separate causally 
disconnected domains of size $H^{-1}$. For each domain, after e-folds $N_{max}-N$,
  \begin{equation}
  \theta_{N_{\rm max}-N-1} = \theta_{N_{\rm max}-N} \pm \delta \theta
  \label{eq:thetapm}
 \end{equation}
In half of the domains the phase evolve towards $\pi$ and in the other half it 
moves towards zero. The process repeats during each e-fold. The size distribution 
of the phase value $\theta$ can be given as a Gaussian,
 \begin{equation}
  P(\bar{\theta}, l) = \frac{1}{\sqrt{2\, \pi} \sigma_l} 
  {\rm exp}\left(- {\frac{(\theta_{N_{\rm max}} - \theta)^2}{2\, \sigma_l}} 
\right),
 \end{equation}
 where
 \begin{equation}
  \sigma_l = \frac{H^2}{4\, \pi^2\, f^2} (N_{\rm max} - N_l).
 \end{equation}

The final stage ends with a collective dynamics of bubble walls as a whole.
The preferred situation is the fluctuations commence with 
$\theta_{N_{\rm max}}< \pi $ so that the present particle horizon contains the 
phase $\theta_{\rm min} =0$. The islands
 with $\theta > \pi $ will move to the final state  $\theta_{\rm min} = 2 \pi$. 
When the size of the domain wall becomes causally connected, it will start to 
self collapse acquiring
spherical form due to self collapse.
The typical size of a closed DW is the correlation length of the last phase 
transition
$L_c = 0.3 M_{\rm Pl}/ \sqrt{g_* \Lambda^2}$, where $g_*$ is the effective 
number
of degrees of freedom in the plasma. 
This relates the characteristic epoch of the Universe with the mass scale of the black holes being formed.
  
To calculate the mass spectrum of PBHs, one has to calculate the size 
distribution of domains that contains phase at the range $\theta > \pi$.
Suppose $V({\bar{\theta}, N})$ is the volume that has been filled with average
phase $\bar{\theta}$ at $N$ e-folds before the end of inflation, then at the
$N-1$ e-folds before the end of inflation,
 \begin{equation}
 V(\bar{\theta}, N-1) = e^3 V(\bar{\theta},N) +
 (V_U(N) - e^3 V(\bar{\theta},N)) P(\bar{\theta}, N-1) \delta \theta,
 \label{eq:size-dist}
\end{equation}
where $V_U \approx e^{3N} H^{-3}$ is the volume of the Universe at N e-fold.
The physical size that leaves the horizon during e-fold number $N(N\le N_{\rm 
max})$ reads,
\begin{equation}
 l = H^{-1} e^N.
\end{equation}
This scale becomes comparable to the FLRW particle
horizon at the moment
\begin{equation}
 t_h = H^{-1} e^{2N}.
\end{equation}
 The amount of energy stored in the vacuum configuration is,
\begin{equation}
 E = \sigma \, t_h^2.
\end{equation}
which gives the value of the black hole mass.  In section \ref{sec:so10} we take up specific application 
of this scenario to the case of $SO(10)$ model with mass scales compatible with this scenario.

\section{$D$-parity induced $SO(10)$ domain walls}
\label{sec:so10}
We begin with a brief recapitulation of \cite{Kibble:1982dd} where the possibility of DW formation in a non-supersymmetric
$SO(10)$ model was considered in detail. 
Consider the symmetry breaking pattern of Spin$(10)$ ,
\begin{equation}
 {\rm Spin(10)} 
 \xrightarrow[54]{M_x} {\rm Spin(6)\times {\rm Spin(4)}} \xrightarrow[126]{M_R} SU(3) \times SU(2) \times U(1) 
 \xrightarrow[10]{M_w} SU(3)\times U(1)
\end{equation}
The scales of symmetry breaking obtained from one-loop renormalization-group analysis are   
\begin{equation}
M_x \simeq 1.9 \times 10^{15}\, {\rm GeV}, \quad M_R \simeq 4.4 \times 10^{13}\, {\rm GeV} .
\end{equation}
The first stage of symmetry breaking is achieved by using the $54$- dimensional scalar field acquiring vacuum
expectation value (vev)  and topologically stable $Z_2$ strings arise.
\begin{equation}
 \langle \phi_{54} \rangle = \phi_0 (2,2,2,2,2,2,-3,-3,-3,-3).
\end{equation}
However, the stability group of this vev contains more elements
than that of $H_0 = SU(6) \times SU(4)$ which can be understood
in the following way. Consider the generators of Spin$(10)$ as 
$ T_{ij} = \sigma_{ij}/2= \left[ \Gamma_i,\Gamma_j\right]/4i, i,j=1,2...10$, where $\Gamma_i$'s are generalised Dirac matrices in $10$ dimension. 
The generators of the subgroup $H_0$ of are then 
$ T_{ab},  1\leq a,b \leq 6$, and $T_{\alpha,\beta}, 7 \leq \alpha \beta \leq 10$.
Consider one element of Spin $(10)$,  $e^{i\theta T_{67}}$
where $\theta$  is the angle of rotation in the $6-7$ plane. Under this transformation, the $6-7$ submatrix of $\langle \phi_{54}\rangle$  transform as,
\begin{equation}
 e^{i \theta T_{67}}
 \begin{pmatrix}
  2 & 0\\
  0 & -3
 \end{pmatrix}
 e^{-i \theta T_{67}} = 
 \begin{pmatrix}
  -\frac{1}{2}+ \frac{5}{2} \cos 2\theta & -\frac{5}{2}\sin 2\theta\\
  -\frac{5}{2}\sin 2\theta & -\frac{1}{2} - \frac{5}{2}\cos 2\theta
 \end{pmatrix}.
 \label{eq:parity-kibble}
\end{equation}
The $\langle \phi_{54} \rangle$ is left invariant for $\theta = n\pi, n\in Z$.

In the second stage of symmetry breaking the $\overline{126}$
develops vacuum expectation value and thereby breaking $H$ down to
$SU(3) \times SU(2) \times U(1)$. A $Z_2$ subgroup of $H_0 =
{\rm Spin(6)\bigotimes {\rm Spin(4)}}$ is also left unbroken
by $\left \langle  \phi_{126} \right \rangle$.  This leads to the formation of $Z_2$ strings \cite{Stern:1985bg}. This transition also breaks
the discrete charge-conjugation symmetry. As a result domain walls are formed that separate the vacua related by charge 
conjugation. The domain walls are topological pseudo-defects 
\cite{Garg:2018trf}, and can be bounded $Z_2$ strings.
 
Assume the $\left \langle  \phi_{126} \right \rangle$ lies 
along the $(\overline{10},1,3)$ direction on the semi-infinite 
plane $y=0^+, x > 0 \, (\varphi =0)$. Then,
\begin{equation}
 \left \langle  \phi_{126} \right \rangle (\varphi)=
 {\rm exp} \left[ \frac{i\varphi}{2} (T_{23} +T_{67}) \right]
 \left \langle  \phi_{126} \right \rangle (\varphi =0), 0\le \varphi \le 2 \pi.
 \label{eq:phi126}
\end{equation}
The above equation means that $\left \langle  \phi_{126} \right \rangle (\varphi =2 \pi)$ lies along the $(10, 1,3)$ direction
which is the charge conjugate of the $(\overline{10},1,3)$
direction. Thus the expectation values of the $\overline{126}$
does not return to its original value after a full rotation around the string.  This leads to the existence of a physical
domain wall along $y = 0, x \ge 0$ semi-infinite plane. The wall
is bounded by the string along z-axis.

These are important illustrative considerations. However the more realistic theory to consider is a supersymmetric one, 
due to the stability of the hierarchy between its high scale and the Standard Model
scale. This is what we take up in the next subsection.

\subsection{The supersymmetric $SO(10)$ GUT}
\label{sec:MSGUT}
We now review the Minimal Supersymmetric Grand Unified Theory  model
within which the scenario of Sec \ref{sec:png} can be realised. One of the major problems of GUTs and specifically SUSY $SO(10)$ is the compatibility
of small neutrino masses and mixings with the high scale of unification    \cite{Aulakh:2008sn,Kopp:2009xt}, \cite{Poh:2017xvg,Raby:2008gh,Dermisek:2006dc,Raby:2003in}. 
However, it is shown in \cite{Aulakh:2013lxa,Babu:2016bmy,Babu:2018tfi,Haba:2020ebr}  that this is achievable
within reasonably economical models, including accommodating inflation in \cite{Aulakh:2012st,Ellis:2016ipm},
still retaining proton decay suppressed. An interesting issue not noted by the proposers of the model
is that due to the presence of $D$-parity within the theory, 
it has degenerate vacua, one of them possessing the Standard Model (SM) $SU(2)_L$ 
as subgroup of its stability group, and the other having instead 
$SU(2)_R$ as a subgroup. This indeed presents the risk of generating dangerous durable domain walls
which can contradict standard cosmology. Here we are interested in only approximate degeneracy of such
vacua, a mechanism that can indeed give rise to an effective potential of the axion type Eq. \eqref{eq:deltaV}.
As for standard cosmology, there are several possibilities that render the vacua non-degenerate, preferring the 
Standard Model as the low energy theory. One possibility for this $SO(10)$ model was considered in \cite{Banerjee:2018hlp},
which produces energy differences sufficient to circumvent the problem to standard cosmology while small enough that the considerations of this section hold. As we shall see the axion of the toy model of Sec. \ref{sec:png} is now a collective degree of freedom connecting symmetry related values 
of some out-of-equilibrium condensates, while the periodic potential in fact arises from the 
effective potential of the Higgs fields, due to the identification of distinct vacua induced
by the $D$-parity. While we study a specific model, the $D$-parity considerations presented here easily generalise to the case of any other supersymmetric $SO(10)$ model with alternative symmetry breaking schemes with relevant Higgs content.

The heavy Higgs part of the renormalisable superpotential of
$SO(10)$ MSGUT is as follows
\cite{Aulakh:2003kg,Bajc:2004xe}:
\be
W =
\frac{m}{4!} \Phi_{ijkl} \Phi_{ijkl} +
\frac{\lambda}{4!} \Phi_{ijkl}\Phi_{klmn}\Phi_{mnij} +
\frac{M}{5!} \Sigma_{ijklm} \bSigma_{ijklm} +
\frac{\eta}{4!} \Phi_{ijkl} \Sigma_{ijmno} \bSigma_{ijmno},
\label{eq:superpotential}
\ee
where $\Phi$ is the $4$-index anti-symmetric representation
$\mathbf{210}$, 
$\Sigma$ is the $5$-index self-dual anti-symmetric representation 
$\mathbf{126}$,
and $\bSigma$ is the $5$-index anti-self-dual anti-symmetric representation 
$\mathbf{\overline{126}}$
of $SO(10)$. In the above expression, all indices range
independently from $1$ to $10$. Below, for brevity of notation, we
shall use $0$ to denote the index $10$ in our expressions.
Of these, the $\Phi$ multiplet is needed for desirable gauge symmetry breaking, 
\begin{equation}
 {\rm Spin(10)} 
 \xrightarrow[210]{M_R} G 
 \end{equation}
where $G$ is generically MSSM for judicious choice of the $210$ vev (vacuum expectation value), however special
values of parameters exist for which $G$ can be one of the intermediate symmetry groups such
as $SU(5)$ or Pati-Salam, or the left-right symmetric group with $U(1)_{B-L}$. For our purpose the vev at the highest scale that emerges towards the end of inflation is relevant. 
The $\bar{\Sigma}$ is needed to give the large mass to the $\nu^c$ state, 
and $\Sigma$ to preserve SUSY while breaking the gauge symmetry.
An additional $10$-plet is desirable for light fermion masses but its required vev is not large and can be ignored for the present purpose. 

The vev's  required to descend to SM gauge group,
expressed in their  Pati-Salam embedding, are 
electric charge conserving \cite{Bajc:2004xe}.
\begin{equation}
\label{eq:bajcvev}
\begin{array}{c}
\Phi_{1234} = \Phi_{1256} = \Phi_{3456} = a, \\
\Phi_{1278} = \Phi_{129\,10} = \Phi_{3478} =
\Phi_{3490} = \Phi_{5678} = \Phi_{5690} = \omega, \\
\Phi_{7890} = p,\\
\\
\Sigma_{a+1,b+3,c+5,d+7,e+9} = i^{a+b+c-d-e} \frac{\sigma}{2^{5/2}},
~~~~ a, b, c, d, e \in \{0,1\}, \\
\\
\bSigma_{a+1,b+3,c+5,d+7,e+9} = i^{-a-b-c+d+e} \frac{\bsigma}{2^{5/2}},
~~~~ a, b, c, d, e \in \{0,1\}.
\end{array}
\end{equation}
The remaining components have zero vevs. The needed energy scale of these vev's is $M_R$ which sets the mass
of the heavy right handed neutrinos upto Yukawa couplings. 

However the theory contains the well known operator, the $D$-parity, flipping these vev's 
to other degenerate vev's which flip $SU(2)_L \leftrightarrow SU(2)_R$, a variant of Eq. \eqref{eq:parity-kibble},
\be
D=\exp(-i\pi J_{23})\exp(i\pi J_{67})
\ee
This model was studied in \cite{Garg:2018trf,Banerjee:2018hlp} where it was shown that 
exactly the same  values of $a$, $\omega$, $p$ ensure F-flatness
at the original vevs as well as at the $D$-flipped vevs. As a result some of the
vev's change sign under $D$-flip as shown in Table \ref{tab:dflip}.
\begin{table}[tbh]
\begin{centering}
\begin{tabular}{| l  c | l | l |}
 \hline
 \hline 
& & & \\
F-term  & \  & Original vev & $D$-Flipped vev \\
& & & \\
\hline 
& & & \\
$
F_{\Phi_{1290}} =
F_{\Phi_{3490}} =
F_{\Phi_{5690}} 
$ 
& = &
$
2 m \omega + 2 \lambda (2 a + p) \omega - \eta \sigma \bsigma 
$
&
$
-2 m \omega - 2 \lambda (2 a + p) \omega + \eta \sigma \bsigma 
$, \\
& & & \\
$
F_{\Phi_{7890}} 
$
& = 
&
$
2 m p + 6 \lambda \omega^2 + \eta \sigma \bsigma 
$
&
$
-2 m p - 6 \lambda \omega^2 - \eta \sigma \bsigma 
$ \\
& & & \\
\hline
\end{tabular}
\end{centering}
\caption{$F$-term vev's transforming non-trivially under $D$-parity flip
}
\label{tab:dflip}
\end{table}
Thus $\mathbf{210}$ has components that can satisfy the criteria of the toy model.

\begin{figure*}[ht]
\begin{center}
\includegraphics[width=0.8\textwidth]{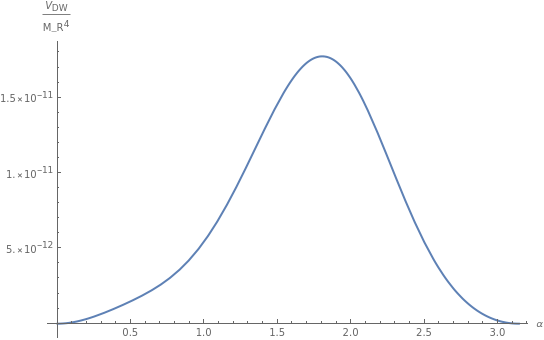}
\caption{The barrier along the $U(1)_D$ trajectory in MSGUT $SO(10)$ model providing an upper bound on actual barrier height, for the case when MSSM can result, modified from Ref. \cite{Garg:2018trf}. $M_R$ is the scale of the vev, $f$ of Sec. \ref{sec:png}. }
\label{fig:barrier-x025}
\end{center}
\end{figure*}

While topologically the same, the two possible subgroups get distinguished in Physics
due to chirality of fermion content. Due to the large parameter space, and the large
number of field components, it is difficult to visualise paths that lead from
one such vacuum to another quasi-degenerate vacuum. However we embed the
transformation $D$ in a $U(1)$ group
\be
U_D(\alpha)=\exp\left\{-i\alpha (J_{23}- J_{67})\right\}
\ee
If we study the effective potential parameterised by $\alpha$ as we traverse from the MSSM vacuum at $\alpha=0$ to $\alpha=\pi$, then the end points are the expected degenerate vacua while the largest energy density along this curve provides an upper bound on the barrier connecting the two \cite{Garg:2018trf}.
The minimal barrier height is difficult to estimate.  In Fig. \ref{fig:barrier-x025} we show a representative barrier recalculated along the lines of Ref. \cite{Garg:2018trf},
a Minimal Supersymmetric Standard Model (MSSM) vacuum and its $D$-parity flipped analogue. The parameters used with reference to Eq. \eqref{eq:superpotential} are $\eta=\lambda=10^{-6}$, and for the case $x=1/4$ of the MSGUT $SO(10)$ model \cite{Aulakh:2003kg}\cite{Bajc:2004xe} so that $m/M=5/9$ \cite{Garg:2018trf}.  
Thus  we can indeed get a barrier much smaller than generic symmetry breaking scale of the theory\footnote{In \cite{Garg:2018trf}, the plots have energy density scale $m^4$ rather than $M_R^4 \sim (m/\lambda)^4$}.

\subsection{PBH creation and comparison with constraints}
\label{sec:constrnt}
\begin{figure*}[ht]
\begin{center}
 \includegraphics[width=0.8\textwidth,height=0.58\textwidth]{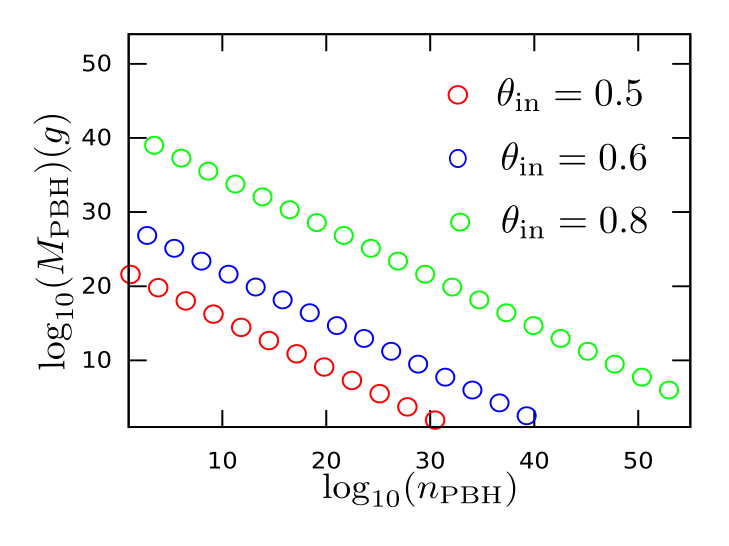}
\vspace{-0.2cm}
\caption{The PBH mass spectrum for different values of the initial phase $\theta_{\rm in}$. Here we have used 
$N_{max}=60$, $H =10^{11} {\rm GeV}, f = 10 H$ and $\Lambda = 10^9{\rm GeV}$.}
\label{fig:pbh}
\end{center}
\end{figure*}

We assume that inflation takes place due to presence of a flat direction in the scalar field potential
of $SO(10)$, or due to an independent inflaton field. The parameter $\alpha$ of $U(1)_D$ of Sec. \ref{sec:MSGUT} is now identified with the axionic degree of freedom $\theta$ of Sec. \ref{sec:png}
and is assumed to follow the equation of motion given in eq. (\ref{eq:eom}). During each 
e-folding the value of $\theta$ changes as given in eq.s (\ref{eq:deltatheta}) 
and (\ref{eq:thetapm}). Here the initial value of $\theta$, $\theta_{N_{\rm max}}\equiv \theta_{\rm in}$ can 
commence with any value which can not be fixed by theory. We take three different values 
$\theta_{\rm in} = 0.5, 0.6$ and $0.8$ to illustrate our results and compare 
with observational constraints. Eventually as the length scale re-enter the horizon after inflation and the fluctuations encoding kink solutions freeze in and DW form.
The surface energy density contributing to the PBH mass is calculated as 
( see for instance \cite{Vilenkin:1981zs})
\begin{equation}
 \sigma = 8f\Lambda^2.
\end{equation}
The one undetermined value $\theta_{\rm in} =\theta_{N_{\rm max}}$ is a free parameter and the results are indeed sensitive 
to the choice of this value. We considered three representative values for it which give interesting results.  
The resulting spectrum of black holes is as shown in Fig. \ref{fig:pbh}. 

\begin{figure*}[ht]
\includegraphics[width=0.49\textwidth,height=0.38\textwidth]{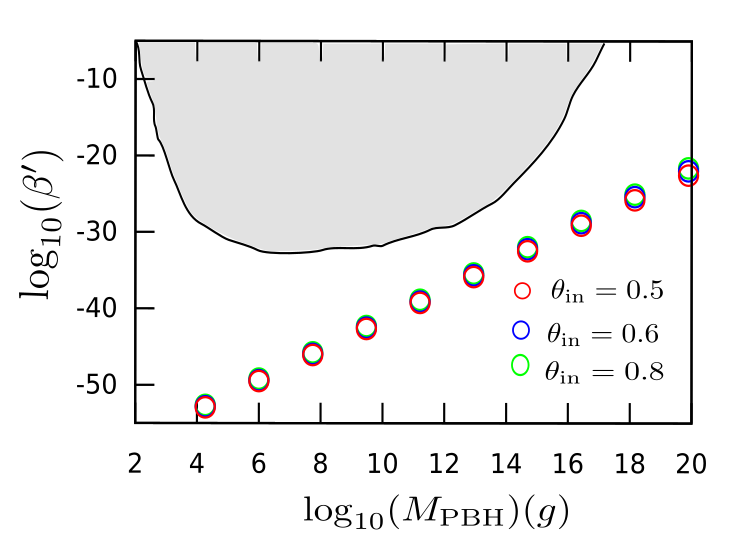}
\hspace{0.158cm}
\includegraphics[width=0.49\textwidth,height=0.38\textwidth]{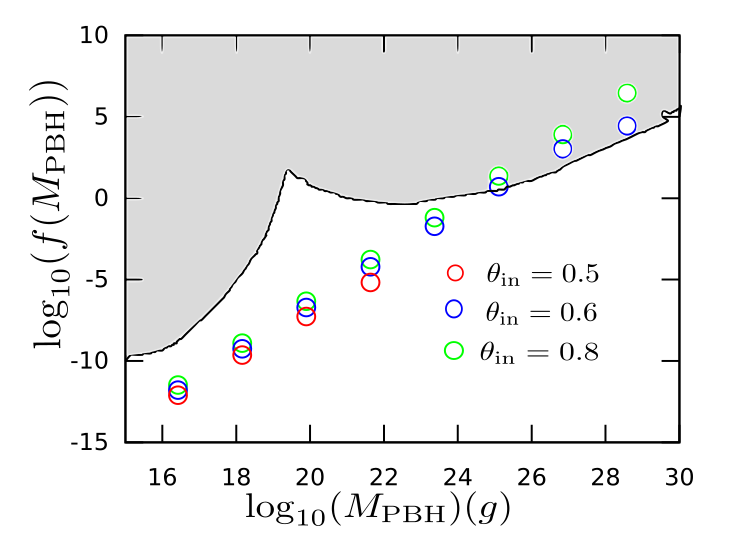} 
\vspace{-0.2cm}
\caption{Evaporation and non-evaporation constraints: in the left panel the shaded area is a rough exclusion region 
sketched corresponding to the Figures 4, 5 and 7 of reference \cite{Carr:2020gox} and corresponds to evaporation constraints.
In the right panel the exclusion region corresponds to the Figure 10 of reference \cite{Carr:2020gox} 
and corresponds to non-evaporation constraints. The different colours correspond to different initial values of 
the phase, $\theta_{\rm in}$. In the vertical axes the quantities $\beta'$ and $f(M_{\rm PBH})$ are calculated 
using  equations (\ref{eq:betapr}) and (\ref{eq:fm}) respectively.}
\label{fig:pbh-constraints}
\end{figure*}

These results must then be transferred to the present epoch and compared with observational constraints.
In appendix \ref{sec:appA} we provide a summary of the constraints on PBH with which we compare our results.
The constraints discussed there are  taken from \cite{Carr:2009jm,Carr:2020gox}  and more detailed references therein. 
In Fig. \ref{fig:pbh-constraints} we present the comparison of our results with the constraints.
For $\theta_{\rm in} = 0.5$, both evaporation and non-evaporation constraints are satisfied, while for $\theta_{\rm in}=0.6$  and $0.8$, the 
evaporation constraint is satisfied  but for the the non-evaporation constraint, the larger mass end of the spectrum can be seen to be already in conflict.
Realistically, the parameter $\theta_{\rm in}$ probably has a randomised distribution and all the values sufficiently close to the 
barrier should be implemented to obtain a comprehensive spectrum of resulting PBH. We expect that such a more detailed study
could constrain the scales $f, H$ and $\Lambda$ as well as demand a better theory for $\theta_{\rm in}$ to confirm avoidance of
late Universe constraints.

\section{Conclusion}
\label{sec:concl}
 As of now we have broad proposals of how early universe fluctuations could have given rise to pre-galactic 
 black holes as discussed for instance in \cite{Carr:2018rid}. A suggestive model has been detailed in \cite{Rubin:2000dq,Rubin:2001yw,Khlopov:2004sc}. 
 In this version,   black holes are created from the collapse of  closed domain walls 
 formed during a second order phase transition which creates topologically non-trivial profiles for the fluctuations
 due to strongly non-equilibrium initial conditions. The non-equilibrium initial conditions are 
generated and also preserved till late epoch by the  inflationary dynamics. The spontaneous 
symmetry breaking freezes in during late stages of inflation. 
Thus the walls which emerge as definite semi-classical objects only at a late epoch are 
preserved from destruction by wall collisions and formation of  holes bounded by strings in them.
 Such walls eventually become causally connected  and begin to contract due to their surface tension.
 This ends  with the formation of black holes with a mass set by  the mass scale  of the wall 
when it enters the horizon.  Thus the size distribution of closed walls gets converted into the mass 
spectrum of resulting primordial black holes.

Among GUTs that can provide a detailed model for such a process, a supersymmetric GUT satisfying proton decay constraints
and stable against its hierarchy with the electroweak scale is an appealing one to consider.
In particular, the $SO(10)$ model accommodates right chiral neutrino states making neutrino masses natural, while also
encompassing an elegant accidental symmetry $D$-parity,  that can exchange left and right chiralities. This discrete symmetry when embedded
in a continuous one-parameter group $U(1)_D$ provides the conditions of the scenario envisaged in \cite{Rubin:2000dq,Rubin:2001yw}.
Here we have worked out the implications to PBH abundance from the mass scales typical of this MSGUT $SO(10)$ based on an earlier
study of the topological pseudo-defects in this theory \cite{Garg:2018trf}. A minimal non-renormalisable extension of the model also ensures 
eventual removal  of the DW as shown in \cite{Banerjee:2018hlp}. Interestingly, the DW which are usually 
considered a nuisance to cosmology are now seen to leave behind relics with macroscopic, stellar scale masses, as a signature of the GUT era. 
The proposed scenario needs the initial conditions on the effective angular variable $\theta$ as an input,  and
 the spectrum of PBH generated depends on the mass scales within the GUT but also on those initial conditions. Thus the study suggests a 
 variety of  conclusions, viz., for the PBH forming scenario, for the GUT and for the statistical distribution of initial fluctuations. As the knowledge 
 of black hole spectrum  improves, one can either constrain the parameters of such GUTs or even rule them out,  or seek refinement of the
mechanisms for such PBH creation.

\section*{Acknowledgement}
UAY acknowledges support from an Institute Chair Professorship. 
 
\appendix
\section{Summary of constraints on PBH}
\label{sec:appA}
The 
PBHs smaller than about $10^{15}g$ would have evaporated
by now and could have many cosmological consequences. PBHs larger 
than $10^{15}g$ are unaffected by Hawking radiation are also
attractive because of the possibility that they provide the 
dark matter content of the Universe. The critical mass for which $\tau$ equals 
the age of the Universe,
 \begin{equation}
  M_* \approx 5 \times 10^{14} ~ g.
 \end{equation}

 For evaporating PBHs the value of the fraction of the Universe in PBHs 
 at the formation time, $t_i$ is related to their number density at $t_i$ and 
$t_0$ is given by,  
 \begin{equation}
  \beta(M) = \frac{ M n_{\rm PBH}}{\rho(T_i)} \approx 
  7.98 \times 10^{-29} \gamma^{-1/2} \left( \frac{g_{*i}}{106.5}\right)^{1/4} 
\times \left( \frac{M}{M_{\odot}} \right)^{3/2} \left( \frac {n_{\rm 
PBH}(t_0)}{1~ {\rm Gpc}^{-3}}\right).
 \end{equation}
 Since $\beta$ always appears in combination with $\gamma^{1/2} g_{*i}^{-1/4} 
h^{-2}$ a new parameter can be defined as 
\begin{equation}
 \beta' (M) \equiv \gamma^{1/2}~ \left( \frac{g_{*i}}{106.75}\right)^{-1/4}
 ~ \left( \frac{h}{0.67}\right)^{-2} \beta(M)
 \label{eq:betapr}
\end{equation}
where $\gamma \sim 0.2$ depends on the details of gravitational collapse. 
The strongest limit comes from $\gamma$-ray, $\beta(10^{15}g)$ $\le 10^{-28}$ 
over all mass ranges. Some of the constraints coming from
possible evaporation of lighter black holes (in the mass range
$10^9$g - $10^{14}$g)  are listed below.

\begin{itemize}
\item The effects of PBH evaporation in CMB for masses smaller than $10^9$ g
\begin{equation}
 \beta' \approx 10^{-5}\left( \frac{M}{10^{9} g} \right)^{-1}, \quad (M< 10^9 
g).
\end{equation}
\item The PBHs in the mass range $10^{11} g < M < 10^{13} g$ produce distortion 
in the CMB spectrum,
\begin{equation}
 \beta' \approx 10^{-16} \left( \frac{M}{10^{11} g} \right)^{-1} \quad (10^{11} 
g < M < 10^{13} g).
\end{equation}
\item The PBHs evaporating after the time of recombination results in small 
scale
anisotropies,
\begin{equation}
 \beta' (M) \approx 3 \times 10^{-30} \left( \frac{f_H}{0.1}\right)^{-1} 
 \left( \frac{M}{10^{13} g} \right)^{3.1} \quad (2.5 \times 10^{13} g \le M \le
 2.4 \times 10^{14} g).
\end{equation}
\item From reionisation of $21$ cm signature 
\begin{equation}
 \beta'(M) \le 3 \times 10^{-29} ~ \left( \frac{M}{10^{14}g} \right)^{7/2}
 \quad M > 10^{14} g.
\end{equation}

\end{itemize} 

Next we list the constraints associated with PBHs which are too large to have 
evaporated
by the present epoch. The current density parameter $\Omega_{\rm PBH}$ 
associated with unevaporated PBHs that form at a redshift $z$ is roughly related 
to $\beta$ by,
\begin{equation}
 \Omega_{\rm PBH} \simeq \beta \Omega_r (1+z) \beta 
 \left( \frac{t}{1~s}\right)^{-1/2} \sim 10^{18} \beta
 \left( \frac{M}{10^5~g} \right)^{-1/2},
\end{equation}
where $\Omega_r \sim 10^{-4}$ is the density parameter of the cosmic microwave 
background. The factor $(1+z)$ arises because the radiation
density scales as $(1+z)^4$ and the PBH density scales as $(1+z)^3$. For 
non-evaporating PBHs the constraints are expressed as constraints
on the fraction of dark matter in the PBHs denoted as $f(M)$,
\begin{equation}
 f(M) = \frac{\Omega_{\rm PBH (M)}}{\Omega_{\rm CDM}} \approx 3.79 \Omega_{\rm 
PBH}
  = 3.81 \times 10^8 \beta' \left( \frac{M}{M_{\bigodot}}\right)^{-1/2}.
  \label{eq:fm}
\end{equation}
The constraints are summarised here.
\begin{itemize}
\item The constraints on PBHs with very low mass come from the femtolensing
of $\gamma$-ray bursts. Assuming the bursts are at a redshift $z\sim 1$, the 
limit is given by,
\begin{equation}
 f(M)<0.1     ( 5 \times 10^{16} g < M < 10^{19} g).
\end{equation}
\item The microlensing observations of stars in Magellanic clouds combining 
MACHO and EROS
results can be given as,
\begin{equation}
 f(M) < 
 \begin{cases}
  1 & 6\times 10^{25}g < M_{\rm PBH} < 3\times 10^{34}g,\\
  0.1 & 10^{27}g < M_{\rm PBH} < 3\times 10^{33}g,\\
  0.04 & 10^{30}g < M_{\rm PBH} < 3\times 10^{32}g.
 \end{cases}
\end{equation}
The OGLE-IV provided stronger limits in the high mass range
\begin{equation}
 f(M) < 
 \begin{cases}
  0.2 & 4\times 10^{32}g < M_{\rm PBH} < 2\times 10^{34}g,\\
  0.09 & 4\times 10^{32}g < M_{\rm PBH} < 1 \times 10^{33}g,\\
  0.06 & 10^{32}g < M_{\rm PBH} < 4 \times 10^{32}g.
 \end{cases}
\end{equation}
\item The studies of the microlensing of quasars were found to exclude all the 
dark matter being in objects with limit $10^{30}g < M_{\rm PBH} < 6\times 
10^{34}g$ and $f(M_{\rm PBH})< 1$.
\item The vulnerability of disruption of binary star systems with wide 
separation with PBHs
constrains the abundance of halo PBHs. The resulting constraints are,
\begin{equation}
 f(M) < 
 \begin{cases}
  (M_{\rm PBH}/ 5\times 10^{33}g)^{(-1)} & 5 \times 10^{35}g < M_{\rm PBH} < 1 
\times 10^{36}g,\\
  0.4 & 1\times 10^{36}g < M_{\rm PBH} < 1 \times 10^{41}g.
 \end{cases}
\end{equation}
\item The survival of globular clusters against tidal disruption by passing PBHs 
gives the limit,
\begin{equation}
 f(M) < 
 \begin{cases}
  (M_{\rm PBH}/3 \times 10^{37}g)^{(-1)} & 3 \times 10^{37}g < M_{\rm PBH} < 1 
\times 10^{39}g,\\
  0.03 & 1\times 10^{39}g < M_{\rm PBH} < 1 \times 10^{44}g.
 \end{cases}
\end{equation}
\item Halo objects can overheat the stars in the Galactic disc and provides the 
limits,
\begin{equation}
 f(M) < 
 \begin{cases}
  (M_{\rm PBH}/3 \times 10^{39}g)^{(-1)} & M_{\rm PBH} < 3 \times 10^{42}g,\\
  0.03 &M_{\rm PBH} >  3 \times 10^{42}g.
 \end{cases}
\end{equation}
\item The survival of galaxies in clusters against tidal disruption by giant 
clusters 
PBHs gives the following limit,
\begin{equation}
 f(M) < 
 \begin{cases}
  (M_{\rm PBH}/7 \times 10^{42}g)^{(-1)} & 7 \times 10^{42}g < M_{\rm PBH} < 1 
\times 10^{44}g,\\
  0.4 & 1\times 10^{44}g < M_{\rm PBH} < 1 \times 10^{47}g.
 \end{cases}
\end{equation}

\item The PBHs can not accrete appreciably in the radiation dominated era. They 
might do so in the period after decoupling that can be analysed through 
Bondi-type analysis. The simultaneous accretion and emission of radiation can 
have significant effect on the thermal history  of the Universe. The emission of 
$X$-rays from accreting PBHs would produce measurable anisotropies and spectral 
distortions on the CMB. The constraints that are be obtained from WMAP data are 
given by,
\begin{equation}
 f(M) < 
 \begin{cases}
  (M_{\rm PBH}/30 \times 10^{33}g)^{(-2)} & 30 \times 10^{33}g < M_{\rm PBH} <   
10^{37}g,\\
  10^{-5} & 10^{37}g \leq M_{\rm PBH} < 10^{44}g, \\
  M_{\rm PBH}/ M_{l=100} & M_{\rm PBH} > 10^{44}g.
 \end{cases}
\end{equation}
where the last expression corresponds to having one PBH on the scale associated
with the CMB anisotropies for $l=100$ modes; $M_{l=100} \simeq 10^{49}$g. 
However the limits associated with accretion of PBHs depend on large number of 
astrophysical parameters and qualitative features such as disc or spherical 
accretion. So the limits may be less secure than the previous ones.

\end{itemize}

%
 \bibliographystyle{JHEP}
\bibliography{pbh,med,SO10inflation}

\providecommand{\href}[2]{#2}\begingroup\raggedright\begin{thebibliography}{10}

\bibitem{LIGOScientific:2016aoc}
{\scshape LIGO Scientific, Virgo} collaboration, B.~P. Abbott et~al.,
  \emph{{Observation of Gravitational Waves from a Binary Black Hole Merger}},
  \href{http://dx.doi.org/10.1103/PhysRevLett.116.061102}{\emph{Phys. Rev.
  Lett.} {\bf 116} (2016) 061102}, [\href{http://arxiv.org/abs/1602.03837}{{\tt
  1602.03837}}].

\bibitem{LIGOScientific:2016sjg}
{\scshape LIGO Scientific, Virgo} collaboration, B.~P. Abbott et~al.,
  \emph{{GW151226: Observation of Gravitational Waves from a 22-Solar-Mass
  Binary Black Hole Coalescence}},
  \href{http://dx.doi.org/10.1103/PhysRevLett.116.241103}{\emph{Phys. Rev.
  Lett.} {\bf 116} (2016) 241103}, [\href{http://arxiv.org/abs/1606.04855}{{\tt
  1606.04855}}].

\bibitem{LIGOScientific:2018mvr}
{\scshape LIGO Scientific, Virgo} collaboration, B.~P. Abbott et~al.,
  \emph{{GWTC-1: A Gravitational-Wave Transient Catalog of Compact Binary
  Mergers Observed by LIGO and Virgo during the First and Second Observing
  Runs}}, \href{http://dx.doi.org/10.1103/PhysRevX.9.031040}{\emph{Phys. Rev.
  X} {\bf 9} (2019) 031040}, [\href{http://arxiv.org/abs/1811.12907}{{\tt
  1811.12907}}].

\bibitem{LIGOScientific:2020ibl}
{\scshape LIGO Scientific, Virgo} collaboration, R.~Abbott et~al.,
  \emph{{GWTC-2: Compact Binary Coalescences Observed by LIGO and Virgo During
  the First Half of the Third Observing Run}},
  \href{http://dx.doi.org/10.1103/PhysRevX.11.021053}{\emph{Phys. Rev. X} {\bf
  11} (2021) 021053}, [\href{http://arxiv.org/abs/2010.14527}{{\tt
  2010.14527}}].

\bibitem{LIGOScientific:2021djp}
{\scshape LIGO Scientific, VIRGO, KAGRA} collaboration, R.~Abbott et~al.,
  \emph{{GWTC-3: Compact Binary Coalescences Observed by LIGO and Virgo During
  the Second Part of the Third Observing Run}},
  \href{http://arxiv.org/abs/2111.03606}{{\tt 2111.03606}}.

\bibitem{Volonteri:2021sfo}
M.~Volonteri, M.~Habouzit and M.~Colpi, \emph{{The origins of massive black
  holes}}, \href{http://dx.doi.org/10.1038/s42254-021-00364-9}{\emph{Nature
  Rev. Phys.} {\bf 3} (2021) 732--743},
  [\href{http://arxiv.org/abs/2110.10175}{{\tt 2110.10175}}].

\bibitem{Caldwell:2022qsj}
R.~Caldwell et~al., \emph{{Detection of Early-Universe Gravitational Wave
  Signatures and Fundamental Physics}},
  \href{http://arxiv.org/abs/2203.07972}{{\tt 2203.07972}}.

\bibitem{Barack:2018yly}
L.~Barack et~al., \emph{{Black holes, gravitational waves and fundamental
  physics: a roadmap}},
  \href{http://dx.doi.org/10.1088/1361-6382/ab0587}{\emph{Class. Quant. Grav.}
  {\bf 36} (2019) 143001}, [\href{http://arxiv.org/abs/1806.05195}{{\tt
  1806.05195}}].

\bibitem{Carr:2018rid}
B.~Carr and J.~Silk, \emph{{Primordial Black Holes as Generators of Cosmic
  Structures}}, \href{http://dx.doi.org/10.1093/mnras/sty1204}{\emph{Mon. Not.
  Roy. Astron. Soc.} {\bf 478} (2018) 3756--3775},
  [\href{http://arxiv.org/abs/1801.00672}{{\tt 1801.00672}}].

\bibitem{Sathyaprakash:2009xs}
B.~S. Sathyaprakash and B.~F. Schutz, \emph{{Physics, Astrophysics and
  Cosmology with Gravitational Waves}},
  \href{http://dx.doi.org/10.12942/lrr-2009-2}{\emph{Living Rev. Rel.} {\bf 12}
  (2009) 2}, [\href{http://arxiv.org/abs/0903.0338}{{\tt 0903.0338}}].

\bibitem{Hawking:1974rv}
S.~W. Hawking, \emph{{Black hole explosions}},
  \href{http://dx.doi.org/10.1038/248030a0}{\emph{Nature} {\bf 248} (1974)
  30--31}.

\bibitem{Carr:1975qj}
B.~J. Carr, \emph{{The Primordial black hole mass spectrum}},
  \href{http://dx.doi.org/10.1086/153853}{\emph{Astrophys. J.} {\bf 201} (1975)
  1--19}.

\bibitem{Ananda:2006af}
K.~N. Ananda, C.~Clarkson and D.~Wands, \emph{{The Cosmological gravitational
  wave background from primordial density perturbations}},
  \href{http://dx.doi.org/10.1103/PhysRevD.75.123518}{\emph{Phys. Rev. D} {\bf
  75} (2007) 123518}, [\href{http://arxiv.org/abs/gr-qc/0612013}{{\tt
  gr-qc/0612013}}].

\bibitem{Jenkins:2020ctp}
A.~C. Jenkins and M.~Sakellariadou, \emph{{Primordial black holes from cusp
  collapse on cosmic strings}},  \href{http://arxiv.org/abs/2006.16249}{{\tt
  2006.16249}}.

\bibitem{Gelmini:2022nim}
G.~B. Gelmini, A.~Simpson and E.~Vitagliano, \emph{{Catastrogenesis: DM, GWs,
  and PBHs from ALP string-wall networks}},
  \href{http://arxiv.org/abs/2207.07126}{{\tt 2207.07126}}.

\bibitem{Kibble:1982dd}
T.~Kibble, G.~Lazarides and Q.~Shafi, \emph{{Walls Bounded by Strings}},
  \href{http://dx.doi.org/10.1103/PhysRevD.26.435}{\emph{Phys. Rev. D} {\bf 26}
  (1982) 435}.

\bibitem{Garg:2018trf}
I.~Garg and U.~A. Yajnik, \emph{{Topological pseudodefects of a supersymmetric
  $SO(10)$ model and cosmology}},
  \href{http://dx.doi.org/10.1103/PhysRevD.98.063523}{\emph{Phys. Rev. D} {\bf
  98} (2018) 063523}, [\href{http://arxiv.org/abs/1802.03915}{{\tt
  1802.03915}}].

\bibitem{Preskill:1992ck}
J.~Preskill and A.~Vilenkin, \emph{{Decay of metastable topological defects}},
  \href{http://dx.doi.org/10.1103/PhysRevD.47.2324}{\emph{Phys. Rev. D} {\bf
  47} (1993) 2324--2342}, [\href{http://arxiv.org/abs/hep-ph/9209210}{{\tt
  hep-ph/9209210}}].

\bibitem{Planck:2018vyg}
{\scshape Planck} collaboration, N.~Aghanim et~al., \emph{{Planck 2018 results.
  VI. Cosmological parameters}},
  \href{http://dx.doi.org/10.1051/0004-6361/201833910}{\emph{Astron.
  Astrophys.} {\bf 641} (2020) A6},
  [\href{http://arxiv.org/abs/1807.06209}{{\tt 1807.06209}}].

\bibitem{Banerjee:2018hlp}
P.~Banerjee and U.~A. Yajnik, \emph{{New ultraviolet operators in
  supersymmetric SO(10) GUT and consistent cosmology}},
  \href{http://dx.doi.org/10.1103/PhysRevD.101.075041}{\emph{Phys. Rev. D} {\bf
  101} (2020) 075041}, [\href{http://arxiv.org/abs/1812.11475}{{\tt
  1812.11475}}].

\bibitem{Rubin:2000dq}
S.~Rubin, M.~Khlopov and A.~Sakharov, \emph{{Primordial black holes from
  nonequilibrium second order phase transition}}, {\emph{Grav. Cosmol.} {\bf 6}
  (2000) 51--58}, [\href{http://arxiv.org/abs/hep-ph/0005271}{{\tt
  hep-ph/0005271}}].

\bibitem{Rubin:2001yw}
S.~G. Rubin, A.~S. Sakharov and M.~Y. Khlopov, \emph{{The Formation of primary
  galactic nuclei during phase transitions in the early universe}},
  \href{http://dx.doi.org/10.1134/1.1385631}{\emph{J. Exp. Theor. Phys.} {\bf
  91} (2001) 921--929}, [\href{http://arxiv.org/abs/hep-ph/0106187}{{\tt
  hep-ph/0106187}}].

\bibitem{Khlopov:2004sc}
M.~Y. Khlopov, S.~G. Rubin and A.~S. Sakharov, \emph{{Primordial structure of
  massive black hole clusters}},
  \href{http://dx.doi.org/10.1016/j.astropartphys.2004.12.002}{\emph{Astropart.
  Phys.} {\bf 23} (2005) 265},
  [\href{http://arxiv.org/abs/astro-ph/0401532}{{\tt astro-ph/0401532}}].

\bibitem{Jeannerot:1995yn}
R.~Jeannerot, \emph{{A Supersymmetric SO(10) model with inflation and cosmic
  strings}}, \href{http://dx.doi.org/10.1103/PhysRevD.53.5426}{\emph{Phys. Rev.
  D} {\bf 53} (1996) 5426--5436},
  [\href{http://arxiv.org/abs/hep-ph/9509365}{{\tt hep-ph/9509365}}].

\bibitem{Kyae:2005vg}
B.~Kyae and Q.~Shafi, \emph{{Inflation with realistic supersymmetric SO(10)}},
  \href{http://dx.doi.org/10.1103/PhysRevD.72.063515}{\emph{Phys. Rev. D} {\bf
  72} (2005) 063515}, [\href{http://arxiv.org/abs/hep-ph/0504044}{{\tt
  hep-ph/0504044}}].

\bibitem{Fukuyama:2008dv}
T.~Fukuyama, N.~Okada and T.~Osaka, \emph{{Realistic Hybrid Inflation in 5D
  Orbifold SO(10) GUT}},
  \href{http://dx.doi.org/10.1088/1475-7516/2008/09/024}{\emph{JCAP} {\bf 09}
  (2008) 024}, [\href{http://arxiv.org/abs/0806.4626}{{\tt 0806.4626}}].

\bibitem{Aulakh:2012st}
C.~S. Aulakh and I.~Garg, \emph{{Supersymmetric Seesaw Inflation}},
  \href{http://dx.doi.org/10.1103/PhysRevD.86.065001}{\emph{Phys. Rev. D} {\bf
  86} (2012) 065001}, [\href{http://arxiv.org/abs/1201.0519}{{\tt 1201.0519}}].

\bibitem{Garg:2015mra}
I.~Garg and S.~Mohanty, \emph{{No scale SUGRA SO(10) derived Starobinsky Model
  of Inflation}},
  \href{http://dx.doi.org/10.1016/j.physletb.2015.10.011}{\emph{Phys. Lett. B}
  {\bf 751} (2015) 7--11}, [\href{http://arxiv.org/abs/1504.07725}{{\tt
  1504.07725}}].

\bibitem{Leontaris:2016jty}
G.~K. Leontaris, N.~Okada and Q.~Shafi, \emph{{Non-minimal quartic inflation in
  supersymmetric $SO(10)$}},
  \href{http://dx.doi.org/10.1016/j.physletb.2016.12.038}{\emph{Phys. Lett. B}
  {\bf 765} (2017) 256--259}, [\href{http://arxiv.org/abs/1611.10196}{{\tt
  1611.10196}}].

\bibitem{Ellis:2016ipm}
J.~Ellis, M.~A.~G. Garcia, N.~Nagata, D.~V. Nanopoulos and K.~A. Olive,
  \emph{{Starobinsky-Like Inflation and Neutrino Masses in a No-Scale SO(10)
  Model}}, \href{http://dx.doi.org/10.1088/1475-7516/2016/11/018}{\emph{JCAP}
  {\bf 11} (2016) 018}, [\href{http://arxiv.org/abs/1609.05849}{{\tt
  1609.05849}}].

\bibitem{Maleknejad:2011jw}
A.~Maleknejad and M.~M. Sheikh-Jabbari, \emph{{Gauge-flation: Inflation From
  Non-Abelian Gauge Fields}},
  \href{http://dx.doi.org/10.1016/j.physletb.2013.05.001}{\emph{Phys. Lett. B}
  {\bf 723} (2013) 224--228}, [\href{http://arxiv.org/abs/1102.1513}{{\tt
  1102.1513}}].

\bibitem{Stern:1985bg}
A.~Stern and U.~A. Yajnik, \emph{{SO(10) Vortices and the Electroweak Phase
  Transition}},
  \href{http://dx.doi.org/10.1016/0550-3213(86)90149-5}{\emph{Nucl. Phys. B}
  {\bf 267} (1986) 158--180}.

\bibitem{Aulakh:2008sn}
C.~S. Aulakh and S.~K. Garg, \emph{{The New Minimal Supersymmetric GUT :
  Spectra, RG analysis and Fermion Fits}},
  \href{http://dx.doi.org/10.1016/j.nuclphysb.2011.12.003}{\emph{Nucl. Phys. B}
  {\bf 857} (2012) 101--142}, [\href{http://arxiv.org/abs/0807.0917}{{\tt
  0807.0917}}].

\bibitem{Kopp:2009xt}
J.~Kopp, M.~Lindner, V.~Niro and T.~E.~J. Underwood, \emph{{On the Consistency
  of Perturbativity and Gauge Coupling Unification}},
  \href{http://dx.doi.org/10.1103/PhysRevD.81.025008}{\emph{Phys. Rev. D} {\bf
  81} (2010) 025008}, [\href{http://arxiv.org/abs/0909.2653}{{\tt 0909.2653}}].

\bibitem{Poh:2017xvg}
Z.~Poh, S.~Raby and Z.-z. Wang, \emph{{Pati-Salam SUSY GUT with Yukawa
  unification}},
  \href{http://dx.doi.org/10.1103/PhysRevD.95.115025}{\emph{Phys. Rev. D} {\bf
  95} (2017) 115025}, [\href{http://arxiv.org/abs/1703.09309}{{\tt
  1703.09309}}].

\bibitem{Raby:2008gh}
S.~Raby, \emph{{SUSY GUT Model Building}},
  \href{http://dx.doi.org/10.1140/epjc/s10052-008-0736-x}{\emph{Eur. Phys. J.
  C} {\bf 59} (2009) 223--247}, [\href{http://arxiv.org/abs/0807.4921}{{\tt
  0807.4921}}].

\bibitem{Dermisek:2006dc}
R.~Dermisek, M.~Harada and S.~Raby, \emph{{SO(10) SUSY GUT for Fermion Masses:
  Lepton Flavor and CP Violation}},
  \href{http://dx.doi.org/10.1103/PhysRevD.74.035011}{\emph{Phys. Rev. D} {\bf
  74} (2006) 035011}, [\href{http://arxiv.org/abs/hep-ph/0606055}{{\tt
  hep-ph/0606055}}].

\bibitem{Raby:2003in}
S.~Raby, \emph{{Phenomenology of the minimal SO(10) SUSY model}},
  \href{http://dx.doi.org/10.1007/BF02705106}{\emph{Pramana} {\bf 62} (2004)
  523--536}, [\href{http://arxiv.org/abs/hep-ph/0304074}{{\tt
  hep-ph/0304074}}].

\bibitem{Aulakh:2013lxa}
C.~S. Aulakh, I.~Garg and C.~K. Khosa, \emph{{Baryon stability on the Higgs
  dissolution edge: threshold corrections and suppression of baryon violation
  in the NMSGUT}},
  \href{http://dx.doi.org/10.1016/j.nuclphysb.2014.03.003}{\emph{Nucl. Phys. B}
  {\bf 882} (2014) 397--449}, [\href{http://arxiv.org/abs/1311.6100}{{\tt
  1311.6100}}].

\bibitem{Babu:2016bmy}
K.~S. Babu, B.~Bajc and S.~Saad, \emph{{Yukawa Sector of Minimal SO(10)
  Unification}}, \href{http://dx.doi.org/10.1007/JHEP02(2017)136}{\emph{JHEP}
  {\bf 02} (2017) 136}, [\href{http://arxiv.org/abs/1612.04329}{{\tt
  1612.04329}}].

\bibitem{Babu:2018tfi}
K.~S. Babu, B.~Bajc and S.~Saad, \emph{{Resurrecting Minimal Yukawa Sector of
  SUSY SO(10)}}, \href{http://dx.doi.org/10.1007/JHEP10(2018)135}{\emph{JHEP}
  {\bf 10} (2018) 135}, [\href{http://arxiv.org/abs/1805.10631}{{\tt
  1805.10631}}].

\bibitem{Haba:2020ebr}
N.~Haba, Y.~Mimura and T.~Yamada, \emph{{Renormalizable $SO (10)$ grand unified
  theory with suppressed dimension-5 proton decays}},
  \href{http://dx.doi.org/10.1093/ptep/ptaa186}{\emph{PTEP} {\bf 2021} (2021)
  023B01}, [\href{http://arxiv.org/abs/2008.05362}{{\tt 2008.05362}}].

\bibitem{Aulakh:2003kg}
C.~S. Aulakh, B.~Bajc, A.~Melfo, G.~Senjanovic and F.~Vissani, \emph{{The
  Minimal supersymmetric grand unified theory}},
  \href{http://dx.doi.org/10.1016/j.physletb.2004.03.031}{\emph{Phys. Lett. B}
  {\bf 588} (2004) 196--202}, [\href{http://arxiv.org/abs/hep-ph/0306242}{{\tt
  hep-ph/0306242}}].

\bibitem{Bajc:2004xe}
B.~Bajc, A.~Melfo, G.~Senjanovic and F.~Vissani, \emph{{The Minimal
  supersymmetric grand unified theory. 1. Symmetry breaking and the particle
  spectrum}}, \href{http://dx.doi.org/10.1103/PhysRevD.70.035007}{\emph{Phys.
  Rev. D} {\bf 70} (2004) 035007},
  [\href{http://arxiv.org/abs/hep-ph/0402122}{{\tt hep-ph/0402122}}].

\bibitem{Vilenkin:1981zs}
A.~Vilenkin, \emph{{Gravitational Field of Vacuum Domain Walls and Strings}},
  \href{http://dx.doi.org/10.1103/PhysRevD.23.852}{\emph{Phys. Rev. D} {\bf 23}
  (1981) 852--857}.

\bibitem{Carr:2009jm}
B.~J. Carr, K.~Kohri, Y.~Sendouda and J.~Yokoyama, \emph{{New cosmological
  constraints on primordial black holes}},
  \href{http://dx.doi.org/10.1103/PhysRevD.81.104019}{\emph{Phys. Rev. D} {\bf
  81} (2010) 104019}, [\href{http://arxiv.org/abs/0912.5297}{{\tt 0912.5297}}].

\bibitem{Carr:2020gox}
B.~Carr, K.~Kohri, Y.~Sendouda and J.~Yokoyama, \emph{{Constraints on
  Primordial Black Holes}},  \href{http://arxiv.org/abs/2002.12778}{{\tt
  2002.12778}}.

\end{thebibliography}\endgroup
\end{document}